# Dimensional Crossover of Mass Anisotropy in Ta-doped $WTe_2$

P. Das[1], D. Das[1], P. Kumar[1], Monika[1] and S. Patnaik[1,*]

[1] *School of Physical Sciences, Jawaharlal Nehru University, New Delhi, India*

[*]*E-mail*: spatnaik@jnu.ac.in

**ABSTRACT:** The observation of extremely large magnetoresistance in $WTe_2$ has attracted considerable attention towards understanding its underlying origin. With its layered van der Waals structure, the question that remains largely unexplored is whether the three-dimensional anisotropic transport characteristics of $WTe_2$ persists under chemical substitution. Here, we present a systematic angle-dependent magneto-transport study of single-crystalline $Ta_xW_{1-x}Te_2$ (x = 0, 0.05, 0.1). The results are analysed within a mass anisotropy scaling framework to extract the mass anisotropy parameter γ as a function of temperature and doping. It is observed that Ta substitution leads to monotonic increase of γ across all temperatures, indicating a progressive deepening of quasi-two-dimensional Fermi surface character. Ta doping also leads to a substantial improvement in crystalline quality, reflected in a pronounced increase in the residual resistivity ratio. Despite weakening electron-hole compensation, the magnetoresistance rises sharply to ~58,211% at x = 0.1, which is assigned to a substantial enhancement in carrier mobility. Rietveld refinement confirms a systematic c-axis contraction with Ta content, identifying the structural origin of the enhanced anisotropy. The mass anisotropy scaling that holds for x = 0 and x = 0.05 breaks down for x = 0.1, where the angular magneto-resistance anisotropy substantially exceeds single-ellipsoid predictions, pointing to a multi-pocket Fermi surface with distinct anisotropies. Nonlinear Hall resistivity provides independent evidence for the underlying multiband character of transport in this system. These findings demonstrate that Fermi surface anisotropy, carrier compensation, and mobility are independent parameters that can lead to tuneable control of large magnetoresistance in topological semimetal $WTe_2$.

**Keywords:** Extremely Large Magnetoresistance (XMR), Mass anisotropy, Fermi surface, 3D scaling relation, Angle dependent magneto-transport.

## I. INTRODUCTION

Materials exhibiting large magnetoresistance (MR), that relates to a substantial change in electrical resistance under an applied magnetic field, are central to spintronic and magnetic sensing technologies [1, 2]. Beyond applications, MR measurements serve as versatile transport probe for extracting carrier density, mobility, and information on scattering mechanisms [3]. In conventional nonmagnetic metals and semiconductors, transverse MR shows quadratic field dependence at low fields that saturates when the mobility-field product approaches unity [4-7]. Non-saturating MR is qualitatively different. It arises most naturally in the semiclassical two-band framework, where nearly compensated electron and hole carriers produce large, unsaturated MR with rich field and temperature dependence [8-10]. This compensation-driven MR is particularly pronounced in nonmagnetic semimetals and is strongly sensitive to both temperature and carrier balance [6, 11-14]. The discovery of extremely large, non-saturating magnetoresistance (XMR) in the layered semimetal $WTe_2$ drew immediate and widespread attention [8, 15]. The XMR, reaching ~$10^5$ % at 2 K and 9 T, was initially attributed to near-perfect electron-hole compensation, supported by ARPES determination of the Fermi surface topology [16]. $WTe_2$ was subsequently identified as a type-II Weyl semimetal, hosting strongly tilted Weyl cones that break Lorentz invariance [17- 19]. Its physical landscape extends further; pressure-induced superconductivity [20], quantum spin Hall insulating behaviour in the monolayer limit [21-23], and high-mobility quantum oscillations in monolayer films [24] have all been reported. Angle-dependent MR studies revealed strongly anisotropic magnetotransport, including an exotic longitudinal linear MR when the field is applied along the tungsten chains [25]. Quantum oscillation measurements identified multiple Fermi pockets, two electron and two hole consistent with a moderately anisotropic three-dimensional Fermi surface [26].

Despite recent progress, the origin of XMR in $WTe_2$ continues to remain as a debatable subject. Charge compensation alone does not fully account for the observations. A parallel line of evidence points to crystal quality and carrier mobility as equally important. Ali et al. showed that $WTe_2$ grown by self-flux has larger mobility and larger MR than crystals grown by chemical vapor transport, directly linking sample quality to XMR [28]. Lv et al. confirmed this systematically: isovalent Mo substitution and Te non-stoichiometry simultaneously reduce mobility and degrade compensation thereby producing a dramatic and monotonic suppression of MR [29, 30]. Barua et al. extended this picture using quantum oscillation measurements on Mo-substituted $WTe_2$, finding that compensation actually improves with Mo doping while MR falls conclusively attributing the MR decrease to reduced transport mobility rather than compensation loss [31]. Flynn et al. compared Mo, Re, and Ta doping in polycrystalline $WTe_2$ and found that aliovalent dopants (Re, Ta) suppress MR far more rapidly than isovalent Mo, attributing this to stronger disruption of the electron-hole balance [32]. However, these studies used polycrystalline samples and low doping levels (≤ 1%), leaving open the question of how single-crystal quality and higher aliovalent Ta concentrations affect transport. What remains unclear is whether compensation, mobility, and Fermi surface anisotropy respond independently to chemical substitution, and which parameter governs the XMR when they evolve in competing directions. A separate but related open question concerns the dimensionality of electronic transport in $WTe_2$. As a van der Waals layered compound, it is often treated as electronically two-dimensional, with the anisotropic MR attributed solely to the field component perpendicular to the ab-plane [8]. However, Thoutam et al. demonstrated through angle-dependent MR measurements that $WTe_2$ is electronically three-dimensional with a small, temperature-dependent mass anisotropy γ, following the scaling $R(H,\theta) = R(\varepsilon_\theta H)$ where $\varepsilon_\theta = (\cos^2\theta + \gamma^{-2}\sin^2\theta)^{1/2}$ and γ is the mass anisotropy [33, 34]. The extracted γ varies from ~1.9 to 5, far smaller than expected for a 2D system. The mass anisotropy tracks the XMR enhancement on cooling and links Fermi surface anisotropy directly to the magnetoresistance [33]. Whether this dimensionality and its temperature dependence are robust to chemical substitution, and whether the scaling framework can probe doping-induced Fermi surface changes, has not been adequately investigated. Beyond addressing dimensionality, angle-dependent MR scaling offers a complementary route to Fermi surface anisotropy information accessible at moderate fields and multiple temperature conditions under which quantum oscillations are typically unresolved in doped systems [31]. This directly captures transport-weighted anisotropy rather than geometric pocket cross-sections, making it particularly suited to chemically substituted systems where disorder may suppress quantum coherence [35].

In this work, we report growth and extensive magnetoresistance characterization of single-crystalline $Ta_xW_{1-x}Te_2$ (x = 0, 0.05, 0.1). Unlike prior Ta-doping studies on polycrystalline samples [32], Ta substitution in single crystals is found to progressively improve crystalline quality. The residual resistivity ratio (RRR) rises monotonically. The angular MR data are analysed within the mass anisotropy scaling framework of Thoutam et al. [33].

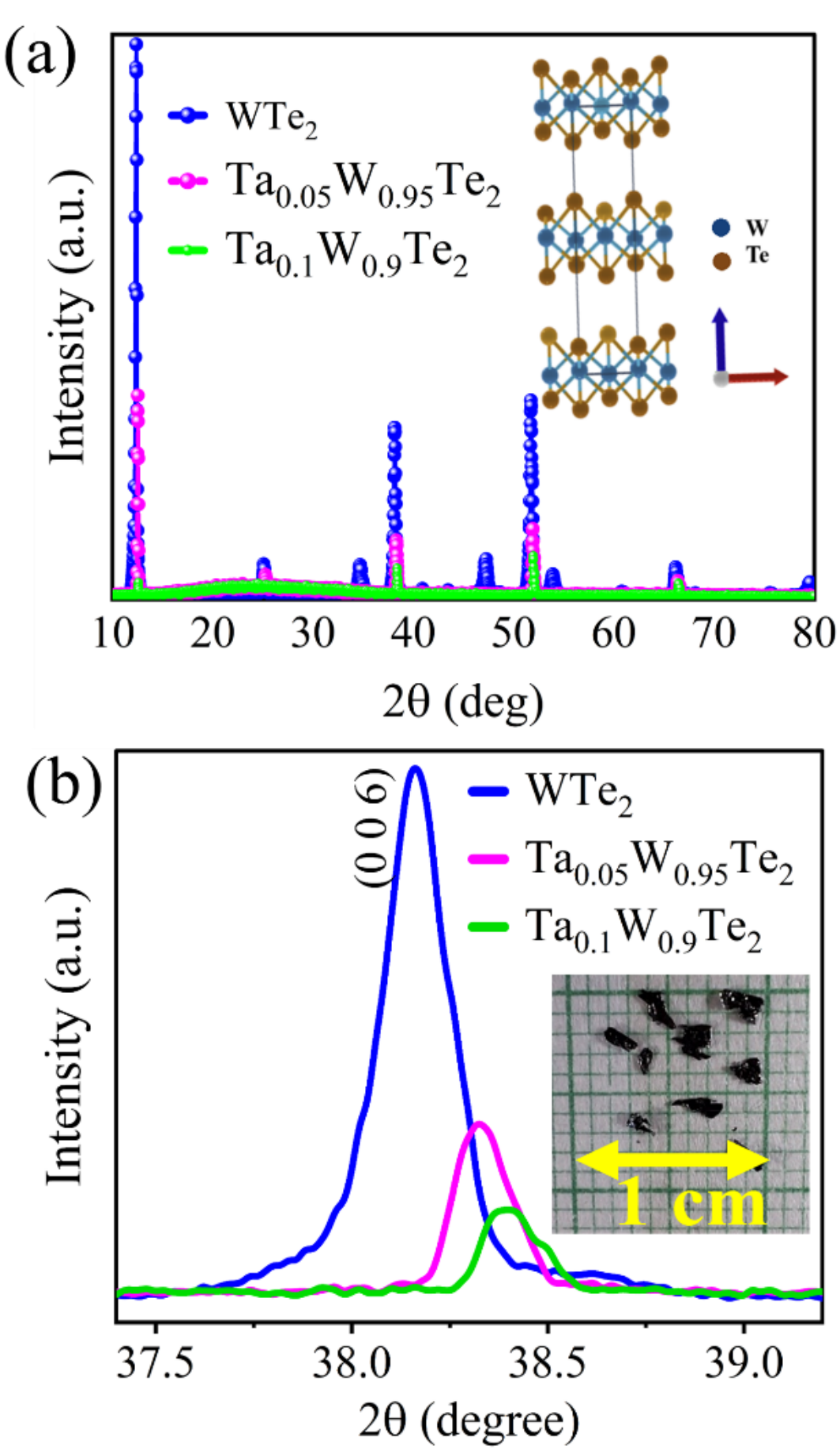

**Figure 1.** (a) Room-temperature X-ray diffraction patterns of single-crystalline $Ta_xW_{1-x}Te_2$ (x = 0, 0.05, 0.1). The inset shows the schematic crystal structure of $WTe_2$. (b) The (0 0 6) Bragg reflection for $Ta_xW_{1-x}Te_2$ (x = 0, 0.05, 0.1), showing a systematic shift toward higher 2θ values with increasing Ta content, indicating progressive c-axis contraction. The inset shows a photograph of representative as-grown single crystals.

## II. EXPERIMENTAL METHODS

Single crystals of Ta-substituted tungsten ditelluride, $Ta_xW_{1-x}Te_2$, with compositions x = 0, 0.05, and 0.10, were grown by the chemical vapor transport (CVT) method with $TeCl_4$ as the transport agent. The stoichiometric amounts of Ta (Alfa Aesar, 99.99%), W (Aldrich, 99.99%), and Te (Aldrich, 99.999%) powders were used as the precursor mixture. The chemicals were ground for 30 minutes to homogenize the mixture. The precursor mixture was loaded into a 12-14 cm long quartz ampoule along with $TeCl_4$ (3 mg/cm$^3$) as the transport agent. The tube was then evacuated, vacuum sealed under the pressure of $10^{-2}$ mbar, and placed in a two-zone furnace with a temperature gradient of 850-750 °C for 20 days, resulting in the formation of plate-like single crystals, deposited in the low-temperature region of the tube. The room-temperature phase and crystal structure were examined using powder X-ray diffraction (XRD) on a Rigaku Miniflex 600 diffractometer having Cu-$K_\alpha$ radiation (λ = 1.54056 Å). Magneto-transport experiments were performed using a *Cryogenic* Cryogen-free superconducting magnet system with fields up to 8 T and temperatures down to 1.6 K. The electrical contacts between the sample and copper wires were made using conductive silver epoxy, which was dried using heat treatment. The standard Four probe method was used for resistivity measurements.

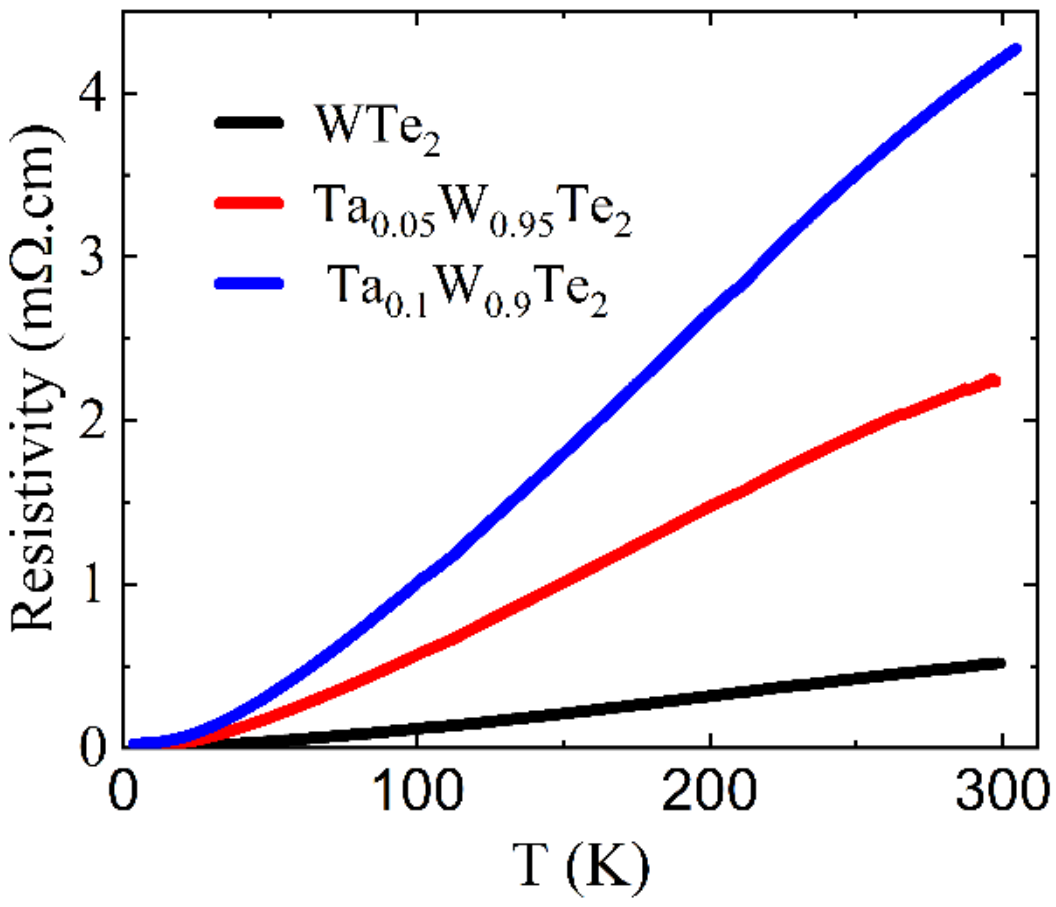

**Figure 2.** Temperature dependence of electrical resistivity ρ(T) for single-crystalline $Ta_xW_{1-x}Te_2$ (x = 0, 0.05, 0.1) in the absence of external magnetic field. All compositions show metallic behaviour. The residual resistivity ratio RRR increases monotonically with Ta substitution (145, 176, and 362 for x = 0, 0.05, and 0.1, respectively), indicating enhanced carrier mobility with doping.

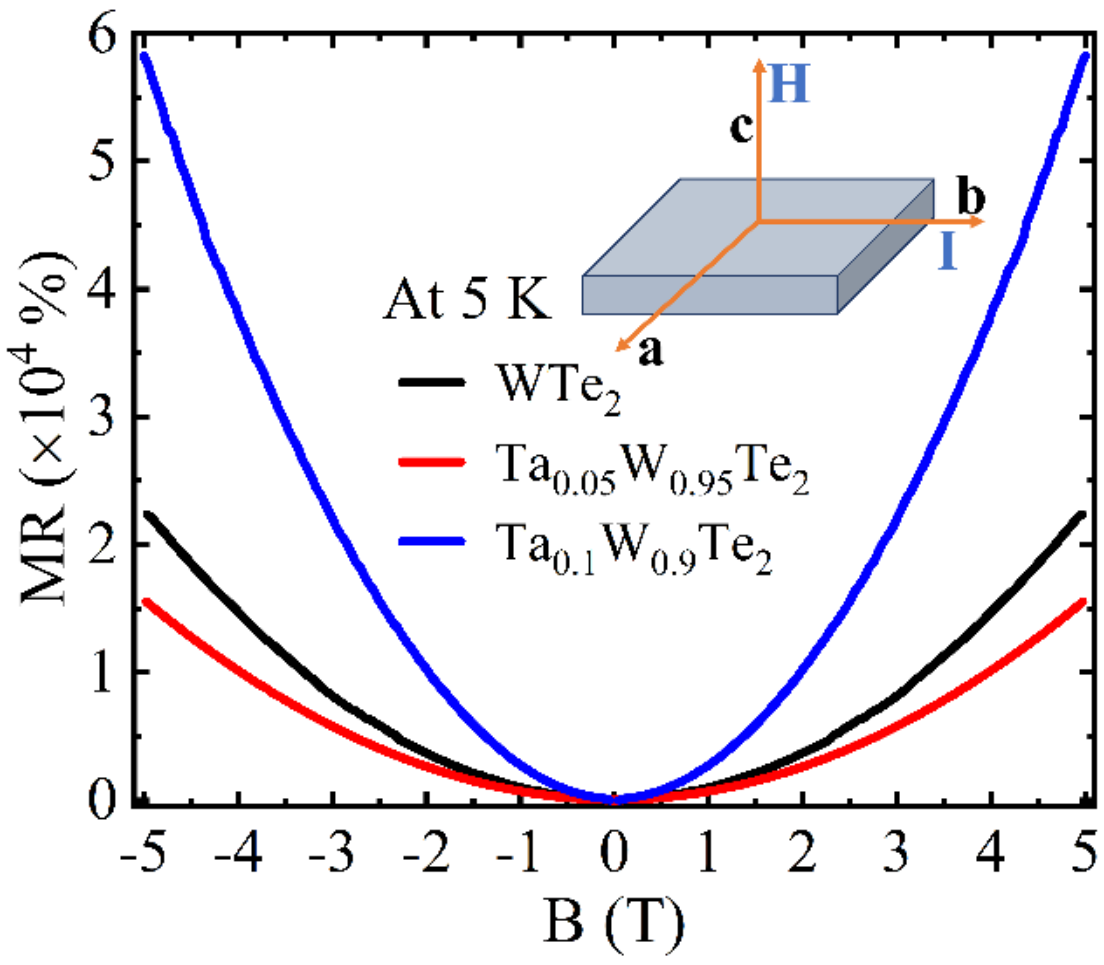

**Figure 3.** Transverse magnetoresistance MR(%) as a function of magnetic field B at 5 K for single-crystalline $Ta_xW_{1-x}Te_2$ (x = 0, 0.05, 0.1), with H applied along the c axis. All compositions display large, non-saturating parabolic MR up to 5 T. The MR evolves non-monotonically with Ta content, decreasing from 22,379% (x = 0) to 15,537% (x = 0.05) before rising sharply to 58,211% (x = 0.1) at 5 T. The inset shows the measurement geometry, with current I flowing along the b axis and field H applied along the c axis.

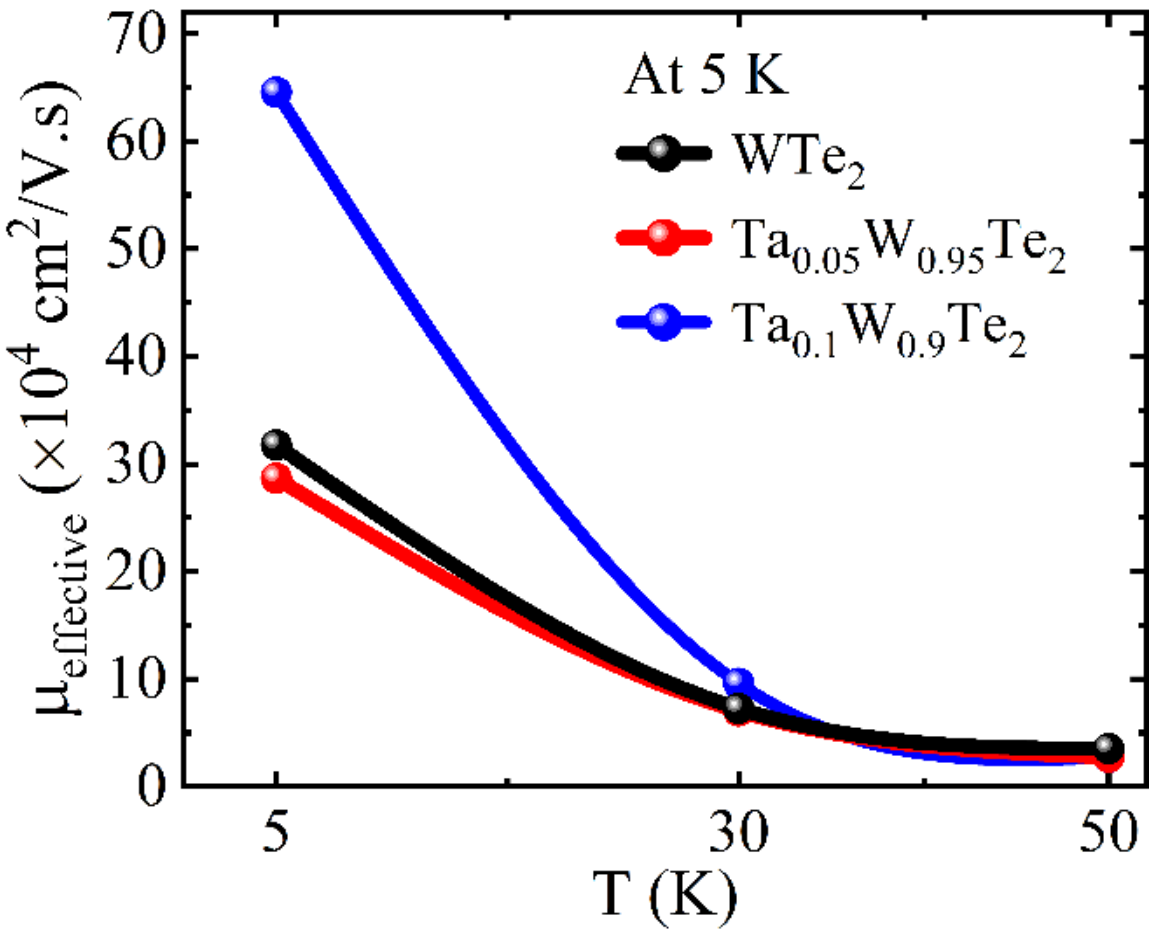

**Figure 4.** Effective carrier mobility $\mu_{eff}$ as a function of temperature for single-crystalline $Ta_xW_{1-x}Te_2$ (x = 0, 0.05, 0.1) at 5 K, 30 K, and 50 K. At 5 K, the mobility of $Ta_{0.1}W_{0.9}Te_2$ (~65 × $10^4$ $cm^2V^{-1}s^{-1}$) is more than double that of pristine $WTe_2$ (~31 × $10^4$ $cm^2V^{-1}s^{-1}$), consistent with the dramatically enhanced MR at x = 0.1.

## III. RESULT AND DISCUSSION

### A. X-ray diffraction

The crystal structure and phase purity of the as-grown $Ta_xW_{1-x}Te_2$ (x = 0, 0.05, 0.10) single crystals were examined by room-temperature powder X-ray diffraction (XRD) on crushed single crystals. $WTe_2$ crystallizes in a distorted octahedral coordination, commonly referred to as the $T_d$ phase, with orthorhombic space group $Pmn2_1$ (No. 31) [36, 37]. Figure 1(a) shows the XRD patterns of all three compositions. The observed reflections are fully indexed to the orthorhombic structure with no additional peaks detected, confirming single-phase purity across the entire doping range. The inset of Figure 1(a) shows the schematic crystal structure of $WTe_2$, and the inset of Figure 1(b) shows a photograph of a representative as-grown single crystal. To resolve the effect of Ta substitution on the lattice, Figure 1(b) shows the (0 0 6) Bragg reflection for all three compositions. A systematic shift toward higher 2θ values is clearly observed with increasing Ta content. By Bragg's law, this indicates a progressive contraction of the c-axis lattice parameter with doping. This is consistent with the substitution of $Ta^{5+}$ (ionic radius ~0.64 Å) at $W^{4+}$ sites (ionic radius ~0.66 Å) [38] the smaller ionic radius of Ta drives a reduction in the interlayer spacing. Rietveld refinement of the powder XRD data confirms the orthorhombic structure with space group $Pmn2_1$ (No. 31) for all compositions. The refined lattice parameters for pristine $WTe_2$ are a = 6.278 Å, b = 3.483 Å, and c = 14.054 Å, in close agreement with reported values [37, 39]. The systematic c-axis contraction with Ta content is consistent with successful incorporation of Ta into the W sites of the lattice. As discussed below, this structural change has a direct consequence on the electronic anisotropy, the reduction in interlayer spacing modifies the effective transport anisotropy parameter γ extracted from the angle-dependent magnetoresistance.

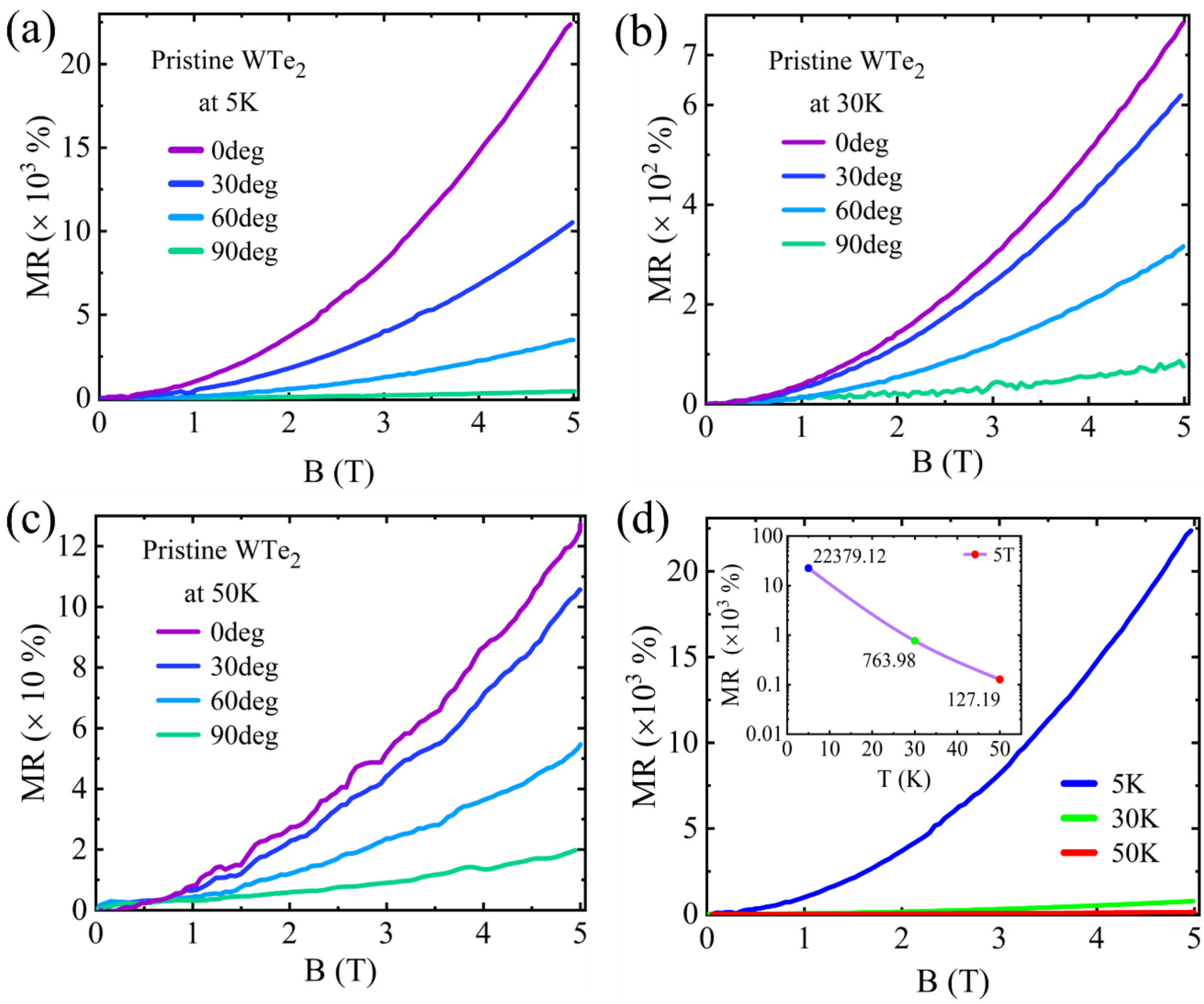

**Figure 5.** Angle-dependent magnetoresistance of pristine $WTe_2$ at (a) 5 K, (b) 30 K, and (c) 50 K for field orientations θ = 0°, 30°, 60°, and 90°, where θ is the angle between the applied field H and the crystallographic c axis. The MR decreases systematically as the field is rotated from H ∥ c (θ = 0°) toward H ∥ ab (θ = 90°), reflecting strong transport anisotropy. The angular anisotropy is most pronounced at 5 K and diminishes with increasing temperature. (d) Transverse MR (H ∥ c) at 5 K, 30 K, and 50 K as a function of field. The inset shows the temperature dependence of MR at 5 T on a logarithmic scale, with values of 22,379%, 764%, and 127% at 5 K, 30 K, and 50 K, respectively.

### B. Temperature-dependent resistivity

Figure 2 shows the temperature dependence of the electrical resistivity ρ(T) for single-crystalline $Ta_xW_{1-x}Te_2$ (x = 0, 0.05, 0.1), measured from 300 K down to 2 K in zero applied magnetic field. All three compositions exhibit metallic behaviour, with resistivity decreasing monotonically on cooling, confirming that the $T_d$ crystal structure of $WTe_2$ is preserved across the entire doping range studied.

The residual resistivity $\rho_0$, extracted from the low-temperature resistivity data, and the residual resistivity ratio RRR = ρ(300 K)/ρ(4 K) are summarized in Table I together with the magnetoresistance values. The pristine $WTe_2$ crystal shows an RRR of ~145, which is higher than typical values reported for CVT-grown crystals [31, 40], reflecting good crystal quality. The RRR increases systematically with Ta substitution, reaching 176 at x = 0.05 and 362 at x = 0.1, a 2.5-fold enhancement over pristine $WTe_2$. This monotonic improvement indicates a progressive reduction in inelastic scattering on cooling, consistent with enhanced carrier mobility in the doped crystals.

The residual resistivity $\rho_0$ evolves non-monotonically with Ta content. It increases slightly from 0.358 μΩ.cm in pristine $WTe_2$ to 0.619 μΩ.cm at x = 0.05, then drops to its lowest value of 0.244 μΩ.cm at x = 0.1. This non-monotonic behaviour is physically significant. At x = 0.05, the slight increase in $\rho_0$ suggests the introduction of additional elastic scattering, consistent with partial disruption of the crystalline order at low doping. At x = 0.1, $\rho_0$ falls below the pristine value, directly confirming that the 10% Ta crystal has the least elastic scattering of all three compositions and hence the highest carrier mobility at low temperature.

**Table I.** Residual resistivity $\rho_0$, RRR, and transverse MR (%) at 5 K and 5 T for single-crystalline $Ta_xW_{1-x}Te_2$ (x = 0, 0.05, 0.1). The RRR increases monotonically with Ta substitution. The residual resistivity $\rho_0$ and MR both evolve non-monotonically, the x = 0.1 composition shows the lowest $\rho_0$ and highest MR, consistent with the highest carrier mobility at this composition.

| COMPOSITION | $\rho_0$ ($\mu\Omega$.cm) | RRR | MR (%) |
|---|---|---|---|
| **$WTe_2$** | 0.358 | 145 | 22379 |
| **$Ta_{0.05}W_{0.95}Te_2$** | 0.619 | 176 | 15537 |
| **$Ta_{0.1}W_{0.9}Te_2$** | 0.244 | 362 | 58211 |

The non-monotonic evolution of $\rho_0$ closely parallels that of the MR itself, providing a direct link between crystal quality and magnetotransport behaviour. This behaviour contrasts sharply with isovalent Mo substitution and Te non-stoichiometry in $WTe_2$, where $\rho_0$ increases monotonically with doping due to progressive defect accumulation [29, 31, 41]. A plausible explanation for the $\rho_0$ reduction at x = 0.1 is that Ta substitution at sufficient concentration suppresses native defects such as Te vacancies common in telluride compounds, or reduces local structural disorder, thereby lowering impurity scattering. This interpretation is consistent with the systematic c-axis contraction observed in the XRD data, which confirms successful incorporation of Ta into the W sites.

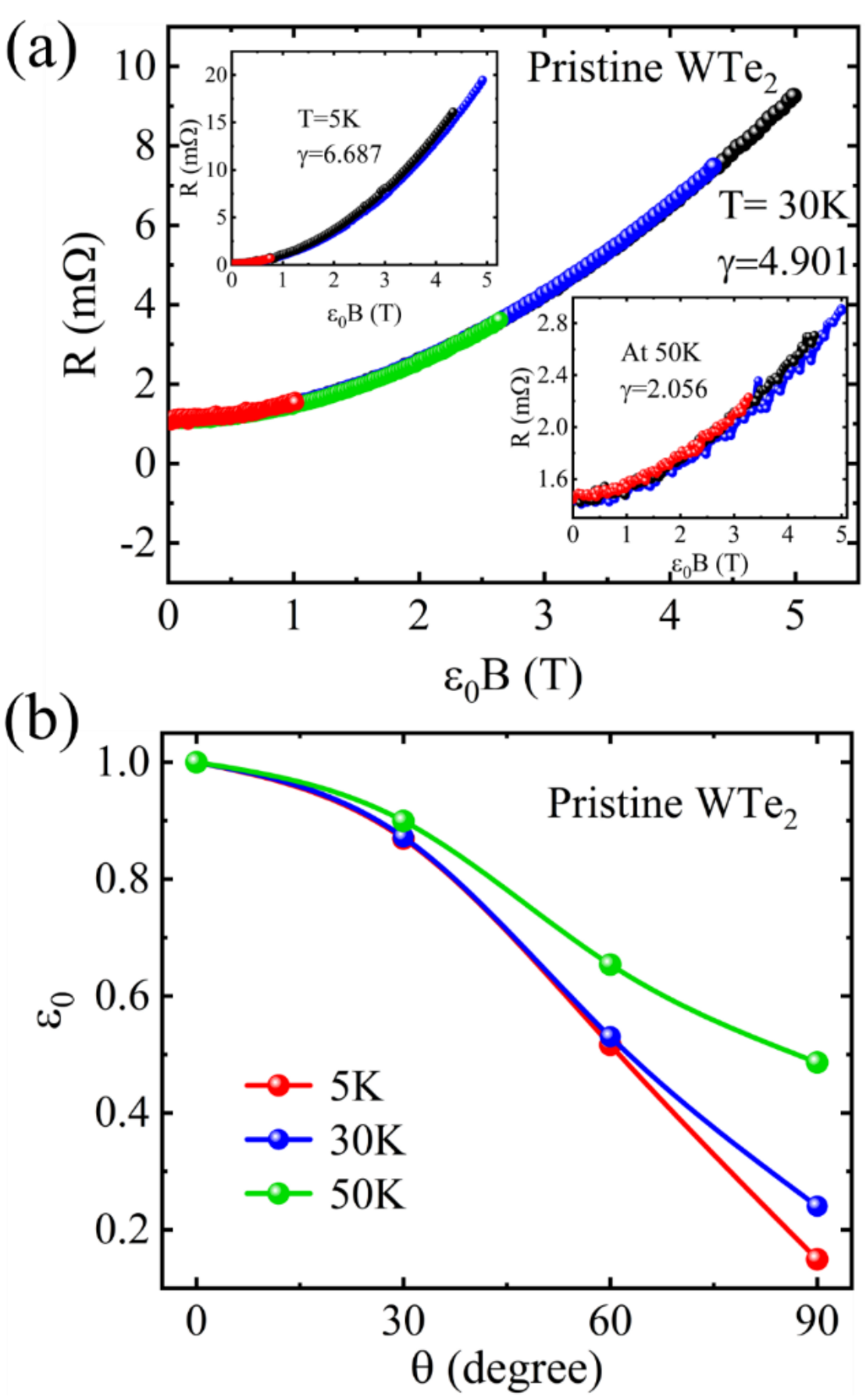

**Figure 6.** Mass anisotropy scaling for pristine $WTe_2$. (a) Resistance R as a function of rescaled field $\varepsilon_0 B$, where $\varepsilon_\theta = (\cos^2\theta + \gamma^{-2}\sin^2\theta)^{1/2}$ at T = 30 K. Resistance curves measured at different field orientations collapse onto a single universal curve, confirming three-dimensional anisotropic transport. The extracted mass anisotropy parameter is γ = 4.901 at 30 K. Insets show the scaling collapse at 5 K (γ = 6.687) and 50 K (γ = 2.056). (b) Angular dependence of the scaling factor $\varepsilon_0$ as a function of field orientation θ at 5 K, 30 K, and 50 K. The scaling factor decreases more steeply at low temperatures, reflecting the stronger anisotropy at low T. Solid lines are fits to $\varepsilon_\theta = (\cos^2\theta + \gamma^{-2}\sin^2\theta)^{1/2}$.

However, to analyse the scattering mechanisms, the resistivity is fitted to:

$$\rho(T) = \rho_0 + AT^2 + BT \quad (1)$$

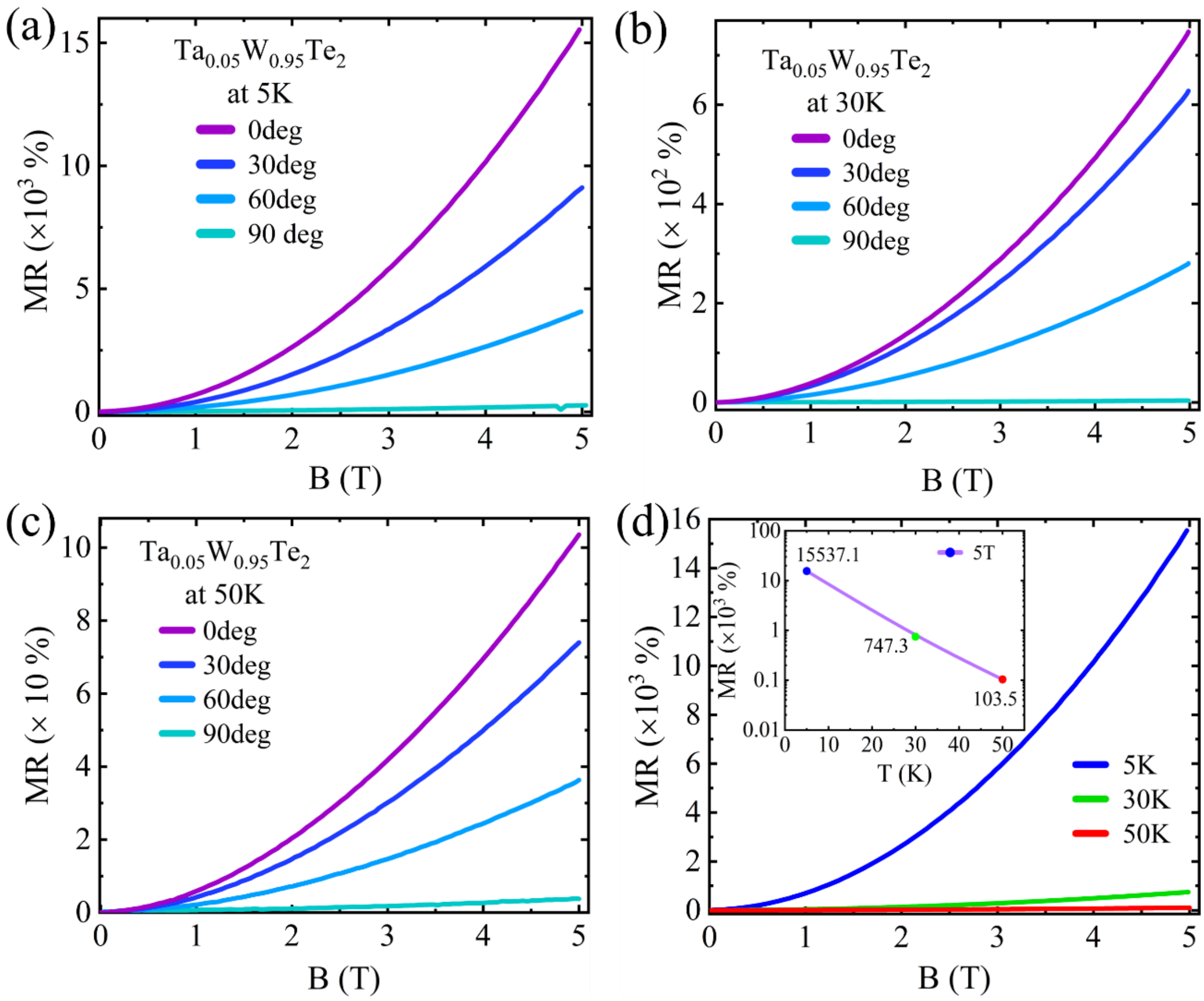

**Figure 7.** Angle-dependent magnetoresistance of $Ta_{0.05}W_{0.95}Te_2$ at (a) 5 K, (b) 30 K, and (c) 50 K for field orientations θ = 0°, 30°, 60°, and 90°, where θ is the angle between H and the c axis. The MR decreases systematically as the field rotates from H ∥ c (θ = 0°) toward H ∥ ab (θ = 90°), consistent with three-dimensional anisotropic transport preserved under moderate Ta doping. The inset of (a) shows the measurement geometry. (d) Transverse MR (H ∥ c) at 5 K, 30 K, and 50 K as a function of field. The inset shows the temperature dependence of MR at 5 T on a logarithmic scale, with values of 15,537%, 747%, and 103% at 5 K, 30 K, and 50 K, respectively.

where $\rho_0$ is the residual resistivity, the $T^2$ term captures electron-electron scattering in the Fermi-liquid regime, and the linear-T term reflects electron-phonon scattering. At low temperatures (T ≤ 50 K), the resistivity of all three compositions is well described by $\rho \propto T^2$, consistent with Fermi-liquid behaviour [42, 34]. At higher temperatures, ρ(T) crosses over to approximately linear behaviour, indicating dominant electron-phonon scattering. The fitted values of $\rho_0$ decrease from x = 0.05 to x = 0.1, in direct correspondence with the RRR enhancement and mobility increase discussed above. The connection between the Fermi-liquid regime identified here and the mass anisotropy parameter γ extracted from angle-dependent magnetoresistance is discussed in the following section.

### C. Transverse magnetoresistance

The transverse magnetoresistance was measured for single-crystalline $Ta_xW_{1-x}Te_2$ (x = 0, 0.05, 0.1) at temperatures from 5 K to 50 K in magnetic fields up to 5 T, with the field applied perpendicular to the sample surface (H ∥ c axis). The magnetoresistance is defined as:

$$MR(\%) = [\rho(B) - \rho(0)]/\rho(0) \times 100\% \qquad (2)$$

where ρ(B) and ρ(0) are the resistivities at applied field B and zero field, respectively. Figure 3 shows the transverse MR at 5 K for all three compositions. A large, non-saturating MR is observed in all compositions up to 5 T. The MR values at 5 K and 5 T are summarized in Table I. The MR evolves non-monotonically with Ta content, it decreases from 22,379% in pristine $WTe_2$ to 15,537% at x = 0.05, then rises sharply to 58,211% at x = 0.1. This non-monotonic behaviour is the central result of this work and is discussed in detail below.

To understand the field dependence of MR, the data are analysed within the semiclassical two-carrier model for a compensated semimetal. The longitudinal resistivity in this model is given by [8, 43]:

$$\rho_{xx} = \frac{(n_e\mu_e+n_h\mu_h)+(n_e\mu_e\mu_h^2+n_h\mu_h\mu_e^2)B^2}{e[(n_e\mu_e+n_h\mu_h)^2+(n_h-n_e)^2\mu_e^2\mu_h^2B^2]} \quad (3)$$

where the electron charge e and $n_e$, $n_h$, $\mu_e$ and $\mu_h$ are total electron concentration, total hole concentration, electron mobility and hole mobility, respectively [8, 43]. In the limit of perfect electron-hole compensation ($n_e = n_h$), this model predicts a non-saturating parabolic MR $\propto \mu_e\mu_h B^2$, consistent with the observed behaviour. The experimental MR follows a power-law dependence MR $\propto B^n$. The extracted exponent n at 5 K decreases monotonically with Ta content: n = 1.980 for $WTe_2$, n = 1.945 for $Ta_{0.05}W_{0.95}Te_2$, and n = 1.899 for $Ta_{0.1}W_{0.9}Te_2$. In a perfectly compensated two-carrier system, n = 2 exactly. The systematic deviation of n below 2 with increasing Ta content therefore signals a progressive degradation of electron-hole compensation across the doping series. Notably, the sample with the largest deviation from n = 2, and hence the worst compensation, simultaneously shows the highest MR. This decoupling of compensation quality from MR magnitude is a key finding of this work, and points to carrier mobility as the dominant parameter controlling XMR at x = 0.1. Figure 4 shows the effective carrier mobility $\mu_{eff}$ as a function of temperature for all three compositions. At 5 K, the effective mobility is ~31 × $10^4$ $cm^2V^{-1}s^{-1}$ for $WTe_2$, ~29 × $10^4$ $cm^2V^{-1}s^{-1}$ for x = 0.05, and ~65 × $10^4$ $cm^2V^{-1}s^{-1}$ for x = 0.1. The mobility at x = 0.1 is more than double that of pristine $WTe_2$ and is comparable to the highest values reported for XMR-exhibiting $WTe_2$ crystals in the literature [33]. With increasing temperature, the mobility of all three compositions decreases rapidly, consistent with the onset of phonon scattering, and the MR falls accordingly. By 50 K, the effective mobilities of all three compositions converge to similar values, explaining why the doping-dependent differences in MR are most pronounced at low temperatures. The non-monotonic MR behaviour can now be understood quantitatively. At x = 0.05, both the mobility (29 × $10^4$ $cm^2V^{-1}s^{-1}$, slightly below pristine) and the compensation (n = 1.945, degraded from 1.980) decrease relative to $WTe_2$. Both factors suppress MR, explaining the drop to 15,537%. At x = 0.1, the compensation degrades further (n = 1.899), but the mobility nearly doubles (65 × $10^4$ $cm^2V^{-1}s^{-1}$). Since MR $\propto \mu_e\mu_h B^2$ in the two-band model, the mobility gain is quadratic in its effect, a factor of ~2 increase in mobility produces a ~4 times increase in MR, overwhelming the compensation penalty and driving the record MR of 58,211%. This mobility-driven enhancement is directly confirmed by the residual resistivity data in Table I, where x = 0.1 shows the lowest $\rho_0$ of all three compositions.

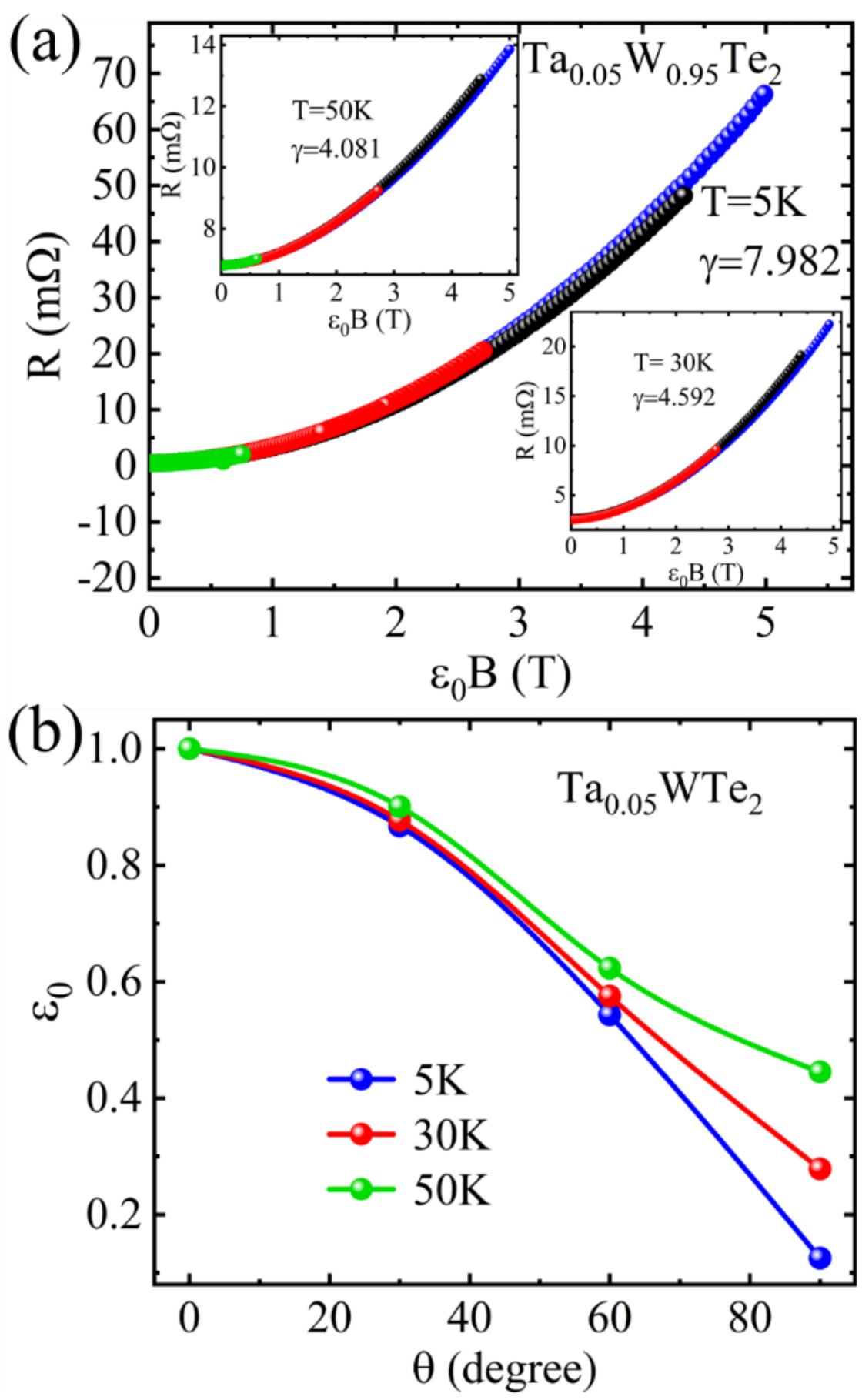

**Figure 8.** Mass anisotropy scaling for $Ta_{0.05}W_{0.95}Te_2$. (a) Resistance R as a function of rescaled field $\varepsilon_0 B$, where $\varepsilon_\theta = (\cos^2\theta + \gamma^{-2}\sin^2\theta)^{1/2}$ at T = 5 K. Resistance curves measured at different field orientations collapse onto a single universal curve, confirming that three-dimensional anisotropic transport is preserved under moderate Ta doping. The extracted mass anisotropy parameter is γ = 7.982 at 5 K. Insets show the scaling collapse at 50 K (γ = 4.081) and 30 K (γ = 4.592). (b) Angular dependence of the scaling factor $\varepsilon_0$ as a function of field orientation θ at 5 K, 30 K, and 50 K. Compared to pristine $WTe_2$, γ is enhanced at all temperatures and shows a weaker temperature dependence, indicating that Ta substitution stabilizes the mass anisotropy against thermal suppression.

### D. Angle-dependent MR study

To probe the dimensionality of electronic transport and the Fermi surface anisotropy, angle-dependent magnetoresistance measurements were performed on all three compositions. The magnetic field B was rotated from θ = 0° (H ∥ c axis, perpendicular to the ab plane) to θ = 90° (H ∥ ab plane), with the current flowing along the a-axis in the ab plane. The angle-dependent MR for all three compositions is shown in Figures 5, 7, and 9. The resistance shows a pronounced anisotropy at all temperatures, with larger values observed when the field is applied closer to the c axis at fixed field magnitude. The angular anisotropy is strongly temperature dependent, being most pronounced at 5 K and diminishing progressively with increasing temperature.

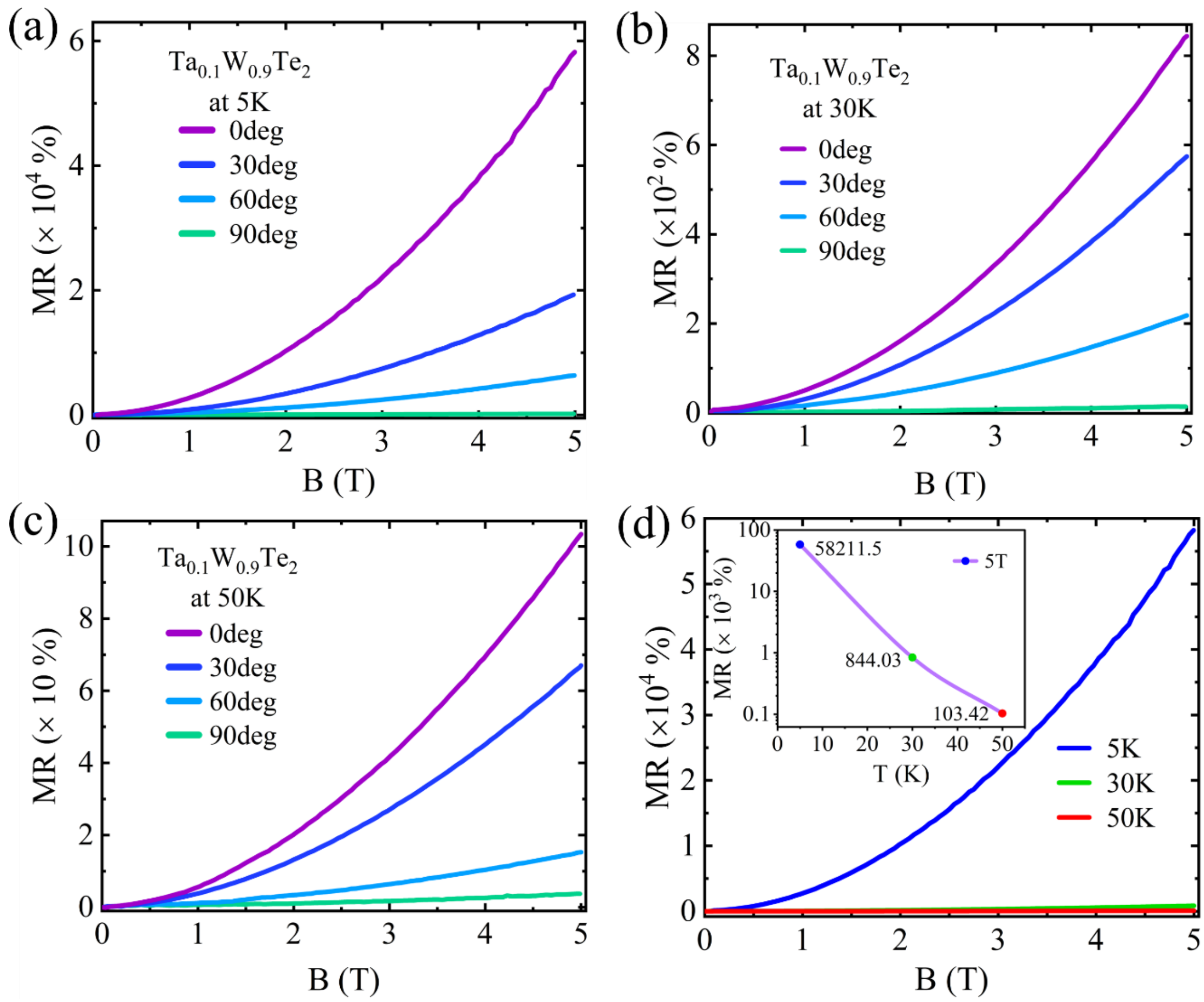

**Figure 9.** Angle-dependent magnetoresistance of $Ta_{0.1}W_{0.9}Te_2$ at (a) 5 K, (b) 30 K, and (c) 50 K for field orientations θ = 0°, 30°, 60°, and 90°, where θ is the angle between H and the c axis. The MR decreases strongly as the field rotates from H ∥ c (θ = 0°) toward H ∥ ab (θ = 90°), with the θ = 90° curve remaining essentially flat at 5 K, indicating the strongest angular anisotropy of all three compositions. The inset of (a) shows the measurement geometry. (d) Transverse MR (H ∥ c) at 5 K, 30 K, and 50 K as a function of field. The inset shows the temperature dependence of MR at 5 T on a logarithmic scale, with values of 58,211%, 844%, and 108% at 5 K, 30 K, and 50 K, respectively. The extremely steep thermal suppression from 5 K to 50 K reflects the fragility of the high-mobility coherent state in this composition.

A key observation that immediately distinguishes $WTe_2$ from a purely two-dimensional system is the presence of significant MR even for in-plane fields (θ = 90°). In a strictly 2D material, H ∥ ab should produce no orbital MR since there is no Lorentz force on carriers confined to the plane. Table II shows the in-plane MR values at 5 T for all three compositions at 5 K, 30 K, and 50 K. Finite in-plane MR is clearly present in all compositions, ruling out a purely 2D description. Notably, the in-plane MR decreases monotonically with Ta content at every temperature, from 420% to 264% to 190% at 5 K directly reflecting the increasing mass anisotropy γ with doping, since larger γ means the in-plane field is rescaled to a smaller effective field.

**TABLE II.** In-plane MR (%) at θ = 90° (H ∥ ab) for $Ta_xW_{1-x}Te_2$ (x = 0, 0.05, 0.1) at 5 T. The in-plane MR decreases monotonically with Ta content at all temperatures, consistent with the progressive increase in mass anisotropy γ. The thermal suppression of in-plane MR is most rapid for x = 0.05, reflecting the thermally stabilized γ in that composition.

| COMPOSITION | MR% (5K) | MR% (30K) | MR% (50K) |
| --- | --- | --- | --- |
| **$WTe_2$** | 420 | 76 | 19 |
| **$Ta_{0.05}W_{0.95}Te_2$** | 264 | 34 | 3.7 |
| **$Ta_{0.1}W_{0.9}Te_2$** | 190 | 14 | 1.6 |

To analyse the anisotropy quantitatively, the angle-dependent resistance data are fitted to the three-dimensional mass anisotropy scaling relation [27, 34]:

$$R(H, \theta) = R(\varepsilon_\theta H) \quad (4)$$

where $\varepsilon_\theta = (\cos^2\theta + \gamma^{-2}\sin^2\theta)^{1/2}$ is the scaling factor that reflects the effective mass anisotropy of the Fermi surface, and $\gamma^2 = m_c/m_{ab}$ is the ratio of effective masses along the c axis and ab plane [33, 44]. In this framework, the resistance curves measured at different field orientations collapse onto a single universal curve when the field is rescaled by $\varepsilon_\theta$. This scaling was originally developed to describe anisotropic superconductors [44] and has been applied theoretically to angle-dependent MR in graphite using a two-band model [45]. For pristine $WTe_2$, the scaling collapse is excellent at all measured temperatures. The extracted anisotropy parameter γ decreases strongly with temperature: γ = 6.687 at 5 K, 4.901 at 30 K, and 2.056 at 50 K. The value at 50 K is significantly smaller than the anisotropy of graphite (~12) [45, 48, 49] and the well-known anisotropic superconductor $YBa_2Cu_3O_7$ (~8) [46], and far below what would be expected for a purely 2D system. This small but finite anisotropy is consistent with quantum oscillation measurements that reveal a 3D Fermi surface of moderate anisotropy in $WTe_2$ [26, 47], and likely reflects strong interlayer coupling arising from distortions of the tellurium layers that accommodate the buckled zigzag tungsten chains. Ta substitution progressively reshapes the Fermi surface anisotropy in two distinct ways. The first is a monotonic enhancement of γ at every temperature, reflecting a deepening of quasi-two-dimensional character as the c-axis contracts with doping. This is physically expected a shorter interlayer spacing weakens the cyclotron orbits increasingly confined to effective hopping along the c axis, making the ab plane and raising the anisotropy of the orbital magnetoresistance response. The second effect is more subtle. In pristine $WTe_2$, γ is strongly temperature dependent, collapsing by 69% between 5 K and 50 K as thermal fluctuations wash out the anisotropic low-temperature electronic structure. Ta doping progressively suppresses this thermal collapse the drop in γ reduces to 49% at x = 0.05 and 43% at x = 0.1. Physically, this stabilization suggests that Ta substitution deepens the quasi-2D potential landscape experienced by carriers, requiring more thermal energy to randomize the anisotropic orbital response. The consequence is directly visible in transport: the angular MR anisotropy ratio MR(0°)/MR(90°) for x = 0.05 remains nearly constant between 5 K and 50 K, whereas in pristine $WTe_2$ it collapses by nearly an order of magnitude over the same range. The single-γ scaling holds well for x = 0 and x = 0.05, confirming that anisotropic 3D transport described by a single ellipsoidal effective mass is preserved under moderate Ta doping. At x = 0.1, this description breaks down. The physical reason is that as γ approaches ~9, the system is near the quasi-2D limit where the Fermi surface can no longer be approximated as a smooth ellipsoid. Different pockets contribute with distinct anisotropies, and no single γ can reconcile the out-of-plane and in-plane responses simultaneously. The observed MR(0°)/MR(90°) ratio of 306 at 5 K is 3.15 times larger than the single-ellipsoid $\gamma^2$ prediction a quantitative measure of how far the actual Fermi surface topology has departed from the simple effective-mass picture. The coincidence of this scaling breakdown with the lowest compensation exponent (n = 1.899) and the highest γ points to a qualitative change in Fermi surface character at x = 0.1, where multiple pockets with distinct anisotropies dominate the low-temperature magnetotransport.

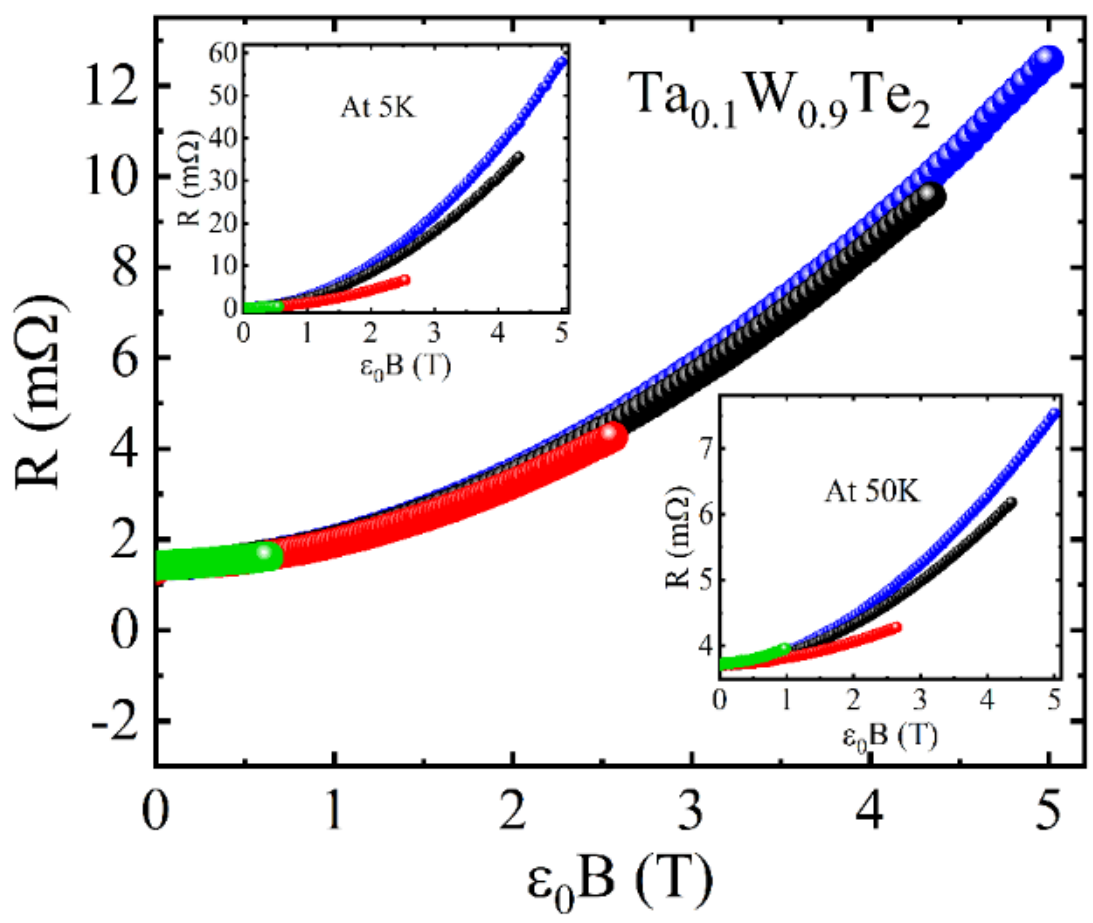

**Figure 10.** Mass anisotropy scaling for $Ta_{0.1}W_{0.9}Te_2$. Resistance R as a function of rescaled field $\varepsilon_0 B$, where $\varepsilon_\theta = (\cos^2\theta + \gamma^{-2}\sin^2\theta)^{1/2}$ at T = 30 K. Unlike pristine $WTe_2$ and $Ta_{0.1}W_{0.9}Te_2$, the resistance curves at different field orientations do not collapse onto a single universal curve, the θ = 0° trace (blue) lies systematically above all other orientations at the same $\varepsilon_0 B$, indicating a breakdown of single-γ anisotropy scaling. Insets show the scaling at 5 K (top left) and 50 K (bottom right), where the collapse failure is most severe at 5 K and improves progressively with temperature. This breakdown points to a multi-pocket Fermi surface with distinct anisotropies in this composition.

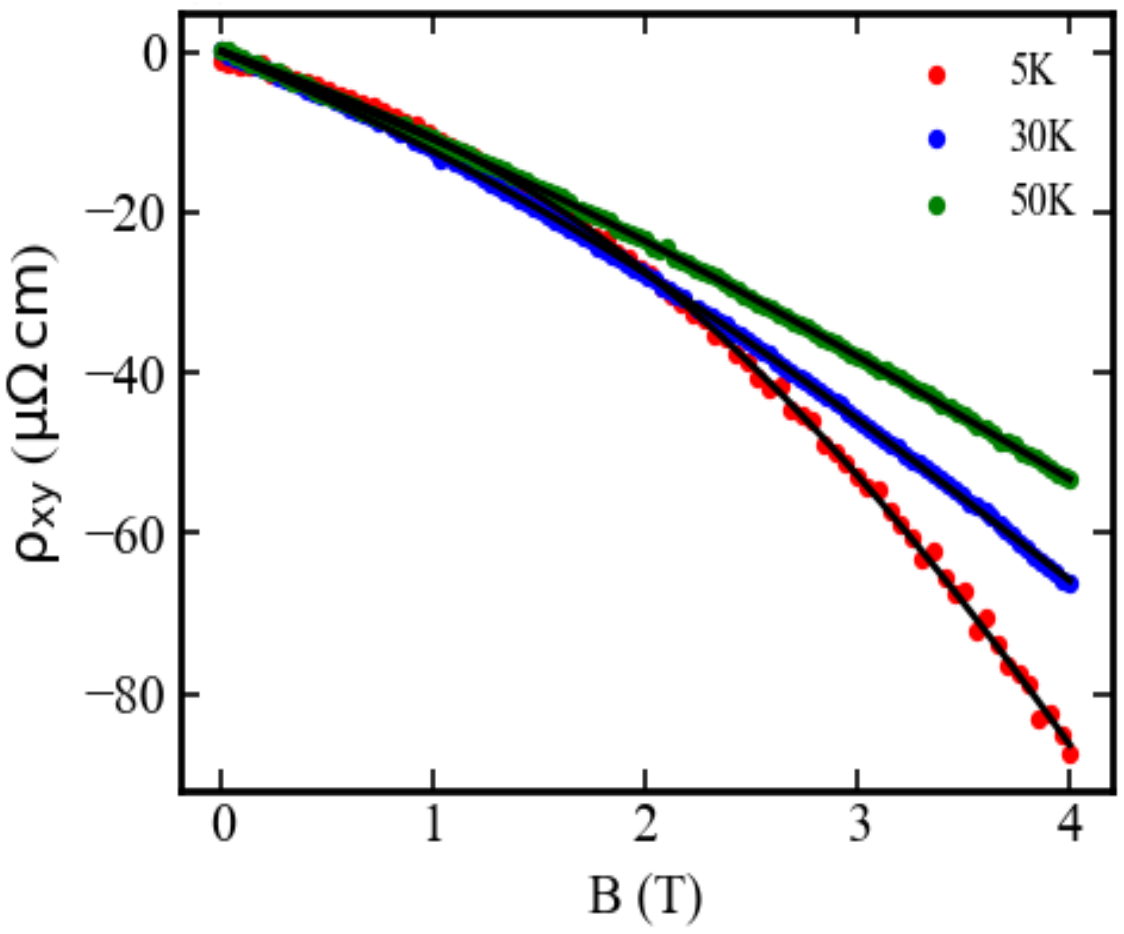

**Figure 11.** Hall resistivity $\rho_{xy}$ as a function of magnetic field B for pristine $WTe_2$ at 5 K, 30 K, and 50 K. The Hall resistivity remains negative over the entire field range, indicating dominant electron-type carriers. At elevated temperatures (30 K and 50 K), $\rho_{xy}$ exhibits approximately linear field dependence, consistent with single-band electron transport. At 5 K, a clear deviation from linearity is observed at high fields, signalling the onset of multiband transport with coexisting electron and hole carriers. Solid lines are fits to the semiclassical two-band model for the Hall resistivity. The nonlinear response at low temperature is consistent with the multiband Fermi surface picture inferred from the anisotropy scaling analysis.

### E. Hall study

Figure 11 shows the field dependence of the Hall resistivity $\rho_{xy}(B)$ for pristine $WTe_2$ at low temperature. The Hall resistivity remains negative over the entire measured field range, indicating that electron-type carriers dominate the transport response. At high temperatures (above ~50 K), $\rho_{xy}(B)$ is approximately linear in field, consistent with single-band electron-dominated transport. With decreasing temperature, $\rho_{xy}(B)$ develops a clear nonlinearity, signalling the onset of multiband transport as both electron and hole carriers become relevant to the Hall response. In systems where electrons and holes contribute simultaneously to conduction, the Hall resistivity is described within the semiclassical two-band model as [8, 43]:

$$\rho_{xy}(B) = \frac{B}{e}\,\frac{(n_h\mu_h^2 - n_e\mu_e^2) + (n_h - n_e)\mu_h^2\mu_e^2B^2}{(n_e\mu_e + n_h\mu_h)^2 + (n_h - n_e)^2\mu_e^2\mu_h^2B^2} \quad (5)$$

where $n_e$ and $n_h$ represent the electron and hole carrier densities, and $\mu_e$ and $\mu_h$ denote their respective mobilities [8, 43]. The nonlinearity of $\rho_{xy}(B)$ at low temperatures arises from the simultaneous contributions of these two carrier types with different mobilities and densities. Analysis of the high-field linear slope of $\rho_{xy}(B)$ yields an electron carrier density of approximately $n_e \sim 3 \times 10^{19}$ $cm^{-3}$, which is typical for semimetallic systems [26]. The deviation from linearity at low fields confirms the coexistence of electron and hole carriers. However, the persistent negative sign of $\rho_{xy}(B)$ at all fields indicates that electron carriers dominate the overall transport response, consistent with the electron-dominated band structure of $WTe_2$ reported by ARPES [16] and quantum oscillation studies [26]. The nearly quadratic MR observed at low temperature is consistent with this high-mobility semiclassical multiband regime. The nonlinear Hall response provides independent evidence for multiband transport in $WTe_2$, the same physics that underlies the breakdown of single-γ anisotropy scaling at x = 0.1 discussed in previous section. In the pristine compound, partial but imperfect carrier compensation, inferred from the n exponent of 1.980 and the persistent negative Hall sign sustains the large XMR through the combined effect of high mobility and near-compensation.

## IV. CONCLUSION

In summary, the angle-dependent magneto-transport properties of single-crystalline $Ta_xW_{1-x}Te_2$ (x = 0, 0.05, 0.1) have been investigated within a mass anisotropy scaling framework. Pristine $WTe_2$ exhibits three-dimensional anisotropic transport with a strongly temperature-dependent mass anisotropy γ. It confirms that its quasi-two-dimensional Fermi surface character is a low-temperature phenomenon. Ta substitution enhances γ monotonically and stabilizes it against thermal suppression, reflecting a progressive deepening of quasi-2D character rooted in c-axis lattice contraction. This is directly confirmed by X-ray diffraction. Crystal quality and carrier mobility improve with Ta content and emerge as the dominant factors behind the observed MR of ~58,211% at x = 0.1. The breakdown of single-γ scaling at x = 0.1, supported independently by power-law exponent analysis, points to a multi-pocket Fermi surface with distinct anisotropies. We conclude that in Ta-doped $WTe_2$, Fermi surface anisotropy, carrier compensation, and mobility are independently-tuneable parameters, each responding distinctly to Ta substitution. This decoupling opens a viable route to engineer large magnetoresistance through selective tuning of individual transport parameters in compensated semimetals and type-II Weyl systems. The mass anisotropy scaling framework proves to be a reliable and accessible probe to decipher Fermi surface characteristics in doped systems that complements conventional transport measurements in regimes where quantum oscillations remain unresolved.

## V. ACKNOWLEDGEMENTS

P. Das and Monika acknowledge UGC-NET SRF for financial support. P. Kumar thank CSIR for providing JRF. We are grateful to the FIST program of the Department of Science and Technology, Government of India for the use of the low-temperature high magnetic field measurement facility at JNU. We acknowledge funding support from DST towards the procurement of chemicals and consumables from the project (DST/NM/TUE/QM10/2109(G)/6) and JNU PAIR network ANRF/PAIR/2025/000029/PAIR.